\documentclass{icrc}

\usepackage{times}
\usepackage{graphicx} % when using Latex and dvips
\begin{document}

\title{Variation of Muon Counts Versus Solar Time}
\author[1]{J. Poirier}
\author[1]{C. D'Andrea}
\affil[1]{Center for Astrophysics at Notre Dame, Physics Dept., University 
of Notre Dame, Notre Dame, Indiana 46556 USA}

\correspondence{John Poirier (JPoirier@nd.edu)}

\runninghead{Poirier: Solar Variations}
\firstpage{1}
\pubyear{2001}

\maketitle

\begin{abstract}
Project GRAND observes 
small variations in the number of incident muons when plotted
versus local solar time.  
%These variations are compared with those of neutron monitor stations 
%at Climax and at Newark.  
The total accumulated number of muons by GRAND over four
years' time is ~140 billion.  The data obtained by Project GRAND are
compared with data from neutron monitor stations 
at Climax and at Newark which detect secondary
neutrons.  
Project GRAND is an array of 64 proportional wire
chamber stations which are  sensitive to secondary muons at energies greater
than 0.1 GeV.  The mean energy of cosmic ray primaries which produce these
muons depends on the spectral index.  For a differential spectral 
index of 2.4, the most
probable gamma ray primary energy is about 10 GeV, the response falling 
steeply below this energy and slowly above this energy \citep{fasso}; 
this most probably energy varies only slowly with spectral index.  
\end{abstract}
%============================
\section{Introduction}
Although cosmic rays are isotropic, variations do exist.  \citet{hall96} have shown that there is a variation expected 
due to co-rotation of the earth and particles in the Interplanetary 
Magnetic Field (IMF).  
As well, this anisotropy should have an amplitude of 0.6\% 
and should be the greatest at 18:00 hours local solar time.  Project GRAND 
uses four years of data to examine this variation with high statistics.  
These data are then compared with data from Climax and Newark 
Neutron Monitors which have a geomagnetic cut-off energy of $\sim$3 GeV..  
%============================
\section{Compton Getting Effect}
A well known anisotropy effect is the Compton Getting effect 
\citep{cge,Cutler}.  
The Compton-Getting effect (CGE) is caused by the motion of the detector relative 
to the rest frame in which the cosmic rays are produced isotropically.  
Equation (1) gives the magnitude of this effect:
\begin{equation}  
\frac{\Delta\alpha(\theta)}{\alpha}=[(2+\gamma)(v_0/c)]\cos\theta
\end {equation}
where $\Delta(\alpha)$/$\alpha$ is the the fractional asymmetry.   
The quantity in square brackets is 
$[(F-A)/A]$ where $F$ is the (forward) counting rate in the direction of 
the velocity 
and $A$ is the average rate; cos$\theta$ gives the projection of the 
secondary cosmic ray along the (forward) direction of $v_o$, 
where $v_o$ is the velocity of the 
detector relative to the production frame of the cosmic rays (where they 
are presumed to be isotropic), $\theta$ is the 
cosmic ray direction relative to $v_0$, and $\gamma$ is the differential 
cosmic ray spectral index describing the energy spectrum of the primary 
cosmic ray. 

Using values of 30 km/s for $v_0$ (the orbital speed of the Earth 
about the sun) and 2.7 for the spectral index,  
Equation (1) gives a CGE amplitude of 4.7$\times10^{-4}$ for the fractional 
forward-backward asymmetry caused by the revolution of the earth around the 
sun.  The orbital speed of the Earth around the sun rather than the velocity of 
the sun in the Galaxy is used since the data will be analyzed in a 
sun-centered frame and the effect of the much larger Galactic speed will 
be cancelled out as the data is averaged over an integer number of years 
(assuming uniform data-taking over this period of time).  
%============================
\section{Acceptance}
Project GRAND is not equally sensitive to all parts of the sky.  It has a 
cutoff angle of 63$^\circ$ from zenith and it, like all ground-based 
detectors, has greater sensitivity to 
cosmic rays coming from near the zenith.  In order to calculate the 
expected asymmetry for Project GRAND it is neccesary to use the correct 
data average value of $\cos \theta$ in Equation (1). 

The acceptance for Project GRAND is given by $Accept$ in 
Equation (2). This equation is 
a combination of three factors, a $\cos \phi$ factor describing the 
projection of muons onto a horizontal surface, a $\cos^2 \phi$ term which 
describes the muon absorption in the Earth's atmosphere because of 
inclination from the vertical direction, and two geometrical $[1-0.537\tan]$ 
factors. The angle $\phi$ is the angle of the muon from zenith.
%============================
\begin{equation}
Accept=[1-0.537\tan \phi_x][1-0.537\tan \phi_y]\cos^3\phi
\end{equation} 
The two geometrical factors of $[1-0.537\tan]$ arise from the use of 
several proportional wire planes stacked above one another.  A muon which 
strikes the top plane but, due to its angle, does not strike the bottom 
plane is not counted.  If a muon is coming in at a completely vertical 
angle, this will not occur while if it is coming in at a shallower angle 
it is more likely not to strike all the planes.  The angle can become so 
shallow that there is no chance for a particle to strike all four planes; that 
limiting projected 
angle is 63$^\circ$ from vertical for Project GRAND.  The 0.537 
in the geometrical factor is the ratio of the x or y width of the detector and 
the vertical spacing between the top and bottom detector.

The combined acceptance function is folded with the $\cos \theta$ term 
from Equation (1) in order to find the asymmetry due to CGE caused by 
orbital velocity that Project GRAND 
expects to see in its counting rate.  A value of 0.922 is obtained for 
$\cos \theta$ and this is multiplied by the CGE amplitude of 
4.7$\times$10$^{-4}$ yielding a predicted effect of 4.3$\times$10$^{-4}$.     
%============================
\section{Experimental Array}
Project GRAND is an array of 64 proportional wire chamber stations. Each 
station contains four pairs of two orthogonal detector planes.  Each plane 
consists of 80 individual detection cells. This arrangement of detectors 
allows for detection of the angle of a particle with a resolution of 0.25 
degrees on a projected plane.  

A steel absorber plate is mounted above the bottom 
pair of planes.  A majority of muons (96\%) do not interact with the 
plate while 96\% of electrons are stopped, scattered, or shower due to the 
plate.  In this way electron tracks can be 
filtered out, keeping only the muon tracks for study.  The information on 
the arrival time and direction of each muon is written onto magnetic 
tape and the data archived for future analysis. 

The magnetic tapes are then analyzed and a file is created containing 
information on the direction of origin of each muon for each complete sidereal 
day.  The information is stored in a $1^\circ\times1^\circ$ grid 
of the number of counts received in right 
ascension vs. declination from 1 to 360 degrees right ascension and 
$-$20 to 90 degrees declination in one degree intervals.
\balance
%============================
\section{Data Analysis}
Data are examined from over four years of running (January 1997 to December 
2000).  In 
that time, data on approximately 111 billion muons were taken.  Each data 
file was tested to prevent spurious variations caused by the operation of 
the experiment itself 
(detectors malfunctioning, for example) from contaminating the data.  A file 
for a given day
was used only if the standard deviation of each degree of right ascension 
divided by the average counts/degree was less than 3\%.  Some 99 billion 
muons passed this test.  

Next the data were converted from right ascension to solar time.  This was 
done by shifting the individual data files in the following manner.  
The values for a given hour of the day come primarily from 
the values of the right ascension which 
were near zenith during that time.  The amount a file's data are shifted 
earlier is proportional to the number of 
days that file is past September 23, the date when the sun is at 12 hr 
in right ascension.  % \citep{aa2000}.  
Data from September 23 are not 
shifted at all, while data from later in the year are shifted 
 earlier by that fraction of 24 hours equal to the fraction of a year which 
 the date is past September 23.  This yields a 
result in terms of local solar time.  

Once 
each day is shifted appropriately, the counts are summed over all declinations.
In order to compare results 
obtained at Project GRAND with those obtained from other detectors, the 
local solar time based on the longitude of the respective experiment was used.    
 
Data from the Climax Neutron Monitor in Climax, Colorado 
($39.4^\circ$N, $116.2^\circ$W) and the 
Newark Neutron Monitor ($39.7^\circ$N, $75.3^\circ$W) are also compared 
(\citet{cliweb}, \citet{newweb}).  Project GRAND is located at 
($41.7^\circ$N, $86.2^\circ$W).
%============================
\section{Fits}
Data from Project GRAND, Climax Neutron Monitor, and Newark Neutron 
Monitor are shown in Figures 1, 2, and 3, respectively.  The data have 
been fit to Equation (1) which describes a curve with a once-per-day and 
a twice-per-day variation, 
%============================
\begin {equation}
y=A+B\cos [15(x-C)] + D\cos [30(x-E)]     
\end {equation}
The $B$ and $D$ parameters are the amplitudes of the once and twice-per-day 
variation, respectively.  The $C$ and $E$ coefficients represent the 
location of the peak in hours of solar time of the once-per-day and twice-per-day variation, 
respectively.  The $A$ parameter is the average value of the counting rate 
for that detector.  The coefficients to Equation (1) are given in Table 1.
%============================
\begin{table}[htbp]
\caption{Paramters of the fit coefficients in Equation (1)}
\begin{center}
\begin{tabular}{cccc}\hline\hline
Parameter & GRAND & Climax & Newark\\
\hline \hline
A & $1.3068\times10^9$ & 4028 & 3400\\
\hline \hline
B & $0.0025\times10^9$ & 11.3 & 9.34\\
\hline
D & $0.0012\times10^9$ & 1.73 & 1.13\\
\hline \hline
C & 16.05 & 13.12 & 12.44\\
\hline
E & 14.71 & 14.08 & 12.27\\
\hline \hline
\end{tabular}
\end{center}
\end{table}

It should be noted that the Climax and Newark data represent the number 
of counts per hour which been been prescaled by a factor of 100 (that is, 
multiply these numbers by $\times$100).  
Project GRAND data represent raw counts per half hour with no scaling.  

\section{Conclusions}
The strength of the once-per-day variation is stronger than the twice-per-day 
variation for all three detectors.  However, for the neutron monitors, 
it is stronger by a factor of seven or eight while for Project GRAND it is 
stronger by only a factor of two.  This would seem to indicate that 
whatever is causing the twice a day variation has a greater effect on 
Project GRAND than on the neutron monitor stations.  One problem in 
attempting to quantify the effect of the variation is that counting rate is 
dependent on the atmosphere's temperature averaged over a column of air 
some 15 km high, a quantity which has little 
relation to ground temperature and about which there are little data.  

According to \citet{hall96} and private communication with \citet{humble} 
a diurnal peak in solar time is caused by cosmic rays which are trapped 
in (or at least partially affected by) 
the interplanetary solar wind, which is co-rotating with the sun.  The 
orbital speed of these particles is about 370 km/sec faster than the 
orbital speed of the Earth, so they overtake the Earth causing an excess of 
counts in that direction.  

The relative strengths of the variations are likely caused by the fact that 
the neutron monitors are less sensitive to the lower energy end of the primary 
spectrum \citep{humble}. The expected maximum component to the anisotropy 
would be at 18  hrs local solar time, in free space.  
However, due to bending from the 
Earth's magnetic field, incoming particles are bent by 15 - 45 degrees so 
the stations' preferred direction of viewing is actually east of their 
physical locations.  Therefore the peak should occur approximately one to 
three hours earlier.   

Project GRAND has, however, a slightly higher primary energy cut-off due 
to the necessity of the primary energy to be above threshold to produce mesons 
which then decay to the muons which are detected by the array; these 
muons must have sufficient energy to 
traverse the $\sim$1300 g/cm$^2$ blanket of air above the detector.
GRAND's higher primary energy cut-off would reduce the magnetic affects 
discussed above and predict smaller hour-of-day 
variations compared to surface neutron detectors which are sensitive 
to lower energies.  

Indeed, comparing the 
once-a-day ratio of $B/A$ values in Table 1, the ratios are: 
GRAND=0.0019, Climax=0.0028, and Newark=0.0027; thus GRAND has only 0.7 
the once-a-day variations of Climax and Newark.  

 \begin{figure*}[t]
 \figbox*{}{}{\includegraphics*[width=11.0cm]{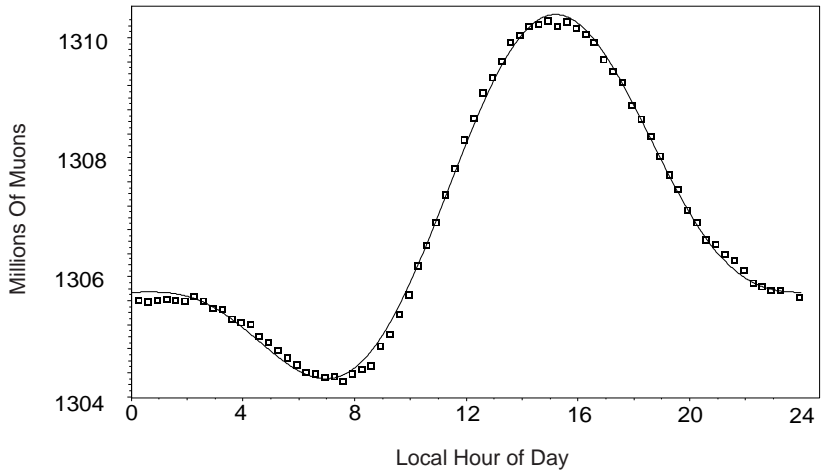}}
 \vspace*{-27.0mm}
 \caption{Millions of muons per 20-minute interval as detected by 
 Project GRAND versus local solar hour (GST-longitude).  
GRAND is located at $41.7^\circ$N,  $86.2^\circ$W.}
 \end{figure*}

 \begin{figure*}[t]
 \figbox*{}{}{\includegraphics*[width=11.0cm]{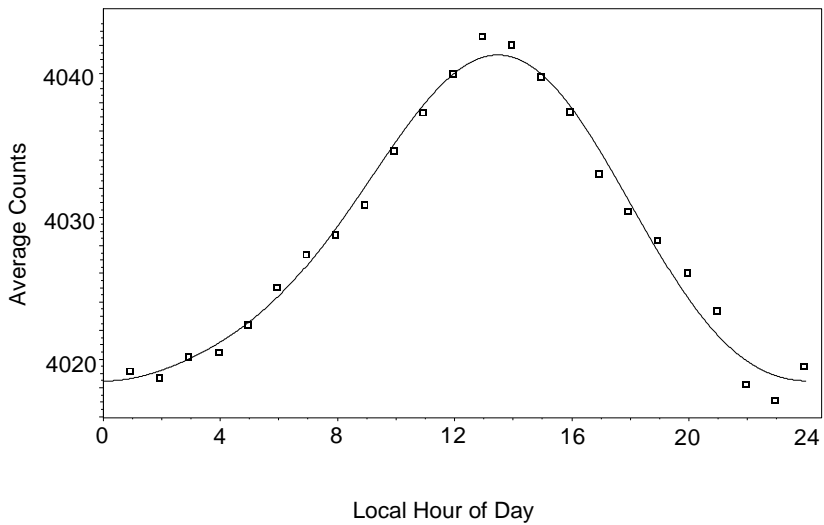}}
 \vspace*{-27.0mm}
 \caption{Neutron counts (prescaled by a factor of 100) 
  per hour for Climax Neutron Monitor vs. local solar 
 time (GST-longitude).  Climax is located at $39.4^\circ$N, $116.2^\circ$W.
%Counts have been prescaled by a factor of 100.}
}
 \end{figure*}

%============================
 \begin{figure*}[t]
 \figbox*{}{}{\includegraphics*[width=11.0cm]{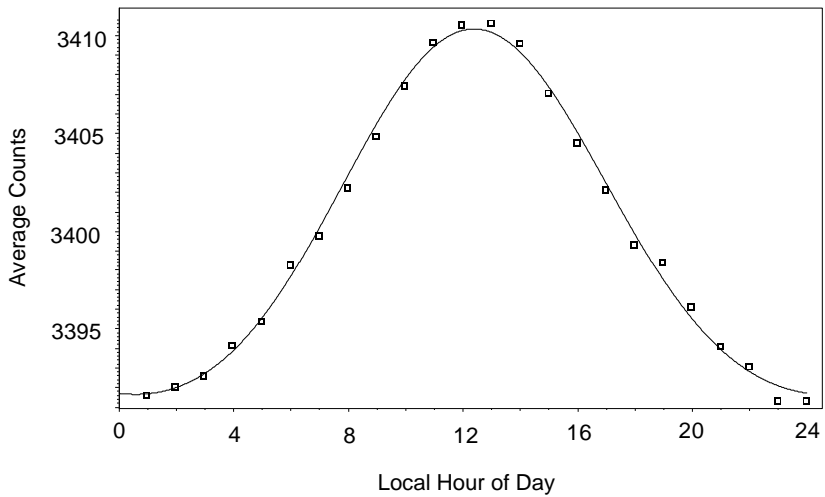}}
 \vspace*{-27.0mm}
 \caption{Neutron counts (prescaled by a factor of 100) 
 per hour for Newark Neutron Monitor vs. local 
 solar time (GST-longitude).  Newark is located at $39.7^\circ$N, 
 $75.3^\circ$W. }
 \end{figure*}
%============================

\begin{acknowledgements}
The authors wish to thank Cliff Lopate (and Climax) and 
Roger Pyle (and Newark) for the use of their data.   
Additional thanks are due John Humble and Jon Vermedahl for 
their assistance.  
Project GRAND is funded through the University of 
Notre Dame and private 
donations.  The Climax Neutron Monitor is operated by the University of 
Chicago and is funded through 
National Science Foundation grant ATM-9912341.  The Newark Neutron Monitor 
is operated by the Bartol Research Institute Neutron Monitor Program and 
is funded through National Science Foundation Gant ATM-0000315.
\end{acknowledgements}
%============================

\end{document}